\documentclass[aps,prb,twocolumn,superscriptaddress,floatfix,longbibliography]{revtex4-2}
\usepackage{graphicx,amsmath,amssymb,bm,xcolor,booktabs}
\usepackage[colorlinks=true,linkcolor=blue,citecolor=blue,urlcolor=blue]{hyperref}
\newcommand{\Nf}{N_{\!f}}
\begin{document}

\title{Quantum skyrmion molecule: two quantum skyrmions bound by coherent magnon exchange}

\author{Ricardo L. Silva}
\email{rmedasilva@gmail.com}
\affiliation{Departamento de Ci\^encias Naturais, Universidade Federal do Esp\'irito Santo, S\~ao Mateus, ES, Brazil}

\date{\today}

\begin{abstract}
Quantum skyrmions in frustrated spin-$1/2$ magnets are bound states of a few flipped spins in a ferromagnetic background. Using exact diagonalization and density-matrix renormalization group calculations for the $J_1$--$J_2$ triangular-lattice model with easy-axis anisotropy $K$, we ask what happens when two such objects are brought together. The binding energy per flipped spin of a single skyrmion has a minimum at $\Nf^{*}=7$ for $K=0.02$--$0.04$, and skyrmions smaller than $\Nf^{*}$ do not exist as pairs: at any distance they coalesce into a single larger object, and pinning wells cannot prevent it. Two skyrmions of natural size, by contrast, form a bound molecule without pinning or external field, with binding energy $U\simeq-0.007J_1$ at $K=0.03$--$0.04$ and an equilibrium separation of $5.5$ lattice spacings that does not depend on $K$ or on the size of the simulation domain; at $K=0.02$, where the single skyrmion is itself marginally bound, the pair is unbound within our resolution. The two constituents keep their identity ($7+7$ flipped spins) and are coupled by a coherent exchange of magnons, visible as number fluctuations between the two halves of the domain in excess of those of two independent skyrmions and as in-plane spin correlations between the two skyrmion walls that reach $0.16$ across the gap. These correlations follow the vortex pattern of the two textures and lock their relative helicity, the quantum counterpart of the helicity-dependent interaction of classical skyrmions. We discuss the molecule as a self-assembled two-qubit element for skyrmion helicity qubits in a single layer.
\end{abstract}

\maketitle

\section{Introduction}
Magnetic skyrmions are localized spin textures with integer topological charge. They have been imaged in chiral magnets and multilayers, where they are stabilized by the Dzyaloshinskii--Moriya interaction, and more recently in centrosymmetric frustrated magnets, where competing exchange interactions play that role~\cite{Nagaosa2013,Fert2017,Okubo2012,Leonov2015,Lin2016,Hayami2016,Kurumaji2019}. In the frustrated case the skyrmion can be as small as a few lattice spacings and carries two internal degrees of freedom that cost no energy classically: its winding $\sigma=\pm1$ (skyrmion or antiskyrmion) and its helicity $\chi$, the angle between the in-plane spin component and the radial direction. Lohani \emph{et al.}~\cite{Lohani2019} showed that in the spin-$1/2$ version of such a magnet the skyrmion survives as a quantum object: a bound state of $\Nf$ flipped spins in the ferromagnetic background, which tunnels between skyrmion and antiskyrmion and whose helicity, being conjugate to $\Nf$, is completely spread in an eigenstate of fixed $\Nf$. Quantum skyrmions have since been studied on lattices, cylinders, and flakes by exact diagonalization, matrix-product states, and neural-network states~\cite{Takashima2016,Sotnikov2021,Haller2022,Siegl2022,Joshi2023,Haller2024,Liscak2026}, in most cases one at a time; where two were considered, they were held at fixed positions by strong pinning potentials~\cite{Haller2024}.

The helicity zero mode is also the basis of the skyrmion qubit proposed by Psaroudaki and Panagopoulos~\cite{Psaroudaki2021,Psaroudaki2022,Psaroudaki2023,Petrovic2025} and developed into a universal gate set by Xia \emph{et al.}~\cite{Xia2023}: a nanoscale skyrmion in a frustrated magnet whose two lowest helicity states encode the qubit. The two-qubit gates proposed so far couple skyrmions in different layers~\cite{Psaroudaki2021,Xia2023} or through a bosonic bus, such as magnons~\cite{Pan2024}; a direct texture--texture gate has been worked out for domain walls~\cite{Zou2023} but not for skyrmions. The most immediate coupling, the interaction between two skyrmions lying side by side in the same film, has not been considered at the quantum level, although classically it is known to depend on the relative helicity of the pair, $U=V(r)+W(r)\cos(\chi_A-\chi_B)$~\cite{Leonov2015,Zhang2017,Capic2020}. Whether this interaction survives quantization, whether it binds the two skyrmions, and whether it correlates their relative helicity are the questions addressed here.

Using exact diagonalization and the density-matrix renormalization group (DMRG) on the model of Ref.~\cite{Lohani2019}, we obtain three results. First, a single quantum skyrmion has a natural size: the binding energy per flipped spin has a minimum at $\Nf^*=7$, independent of the anisotropy in the range studied. Second, skyrmions smaller than $\Nf^*$ do not exist as pairs; at any distance they coalesce into one larger object, and pinning cannot prevent it. Third, two skyrmions of natural size form a bound state with no pinning and no external field, with binding energy $\sim0.007J_1$ and equilibrium separation of $5.5$ lattice spacings, held together by a coherent exchange of magnons that correlates the in-plane spin components of the two walls and locks their relative helicity. We call this bound state a \emph{quantum skyrmion molecule}: the two constituents keep their identity and are held by a shallow well, in the sense in which two atoms form a van der Waals dimer. The molecule is therefore a pair of helicity qubits whose coupling is built into their own interaction.

The paper is organized as follows. Section~\ref{sec:model} defines the model, the finite flakes, and the observables; Sec.~\ref{sec:single} establishes the natural size; Sec.~\ref{sec:coal} shows the coalescence below it; Sec.~\ref{sec:mol} presents the molecule and the tests of its binding energy; Sec.~\ref{sec:hel} analyzes the magnon-exchange coupling; Sec.~\ref{sec:disc} discusses energy scales, implications, and open points. Two appendices collect the convergence tests and the entanglement measures.

\section{Model and methods}
\label{sec:model}

\subsection{Hamiltonian and the frozen-background flake}
We study the spin-$1/2$ model of Ref.~\cite{Lohani2019} on the triangular lattice,
\begin{equation}
H=-J_1\!\sum_{\langle ij\rangle}\!\bm S_i\!\cdot\!\bm S_j+J_2\!\!\sum_{\langle\langle ij\rangle\rangle}\!\!\bm S_i\!\cdot\!\bm S_j-K\!\sum_{\langle ij\rangle}\!S_i^zS_j^z-B\sum_i S_i^z,
\label{eq:H}
\end{equation}
with ferromagnetic nearest-neighbor exchange $J_1=1$, antiferromagnetic next-nearest-neighbor exchange $J_2=0.5$, which frustrates the ferromagnet and makes its spin stiffness negative, and an easy-axis anisotropy $K>0$ that keeps flipped spins together. The model conserves $S^z_{\rm tot}$, so the number of flipped spins with respect to the fully polarized state, $\Nf=S^z_{\rm fm}-S^z_{\rm tot}$, is a good quantum number and all calculations are done in a sector of fixed $\Nf$. In such a sector the Zeeman term is a constant, and we set $B=0$ throughout. A quantum skyrmion is a bound state of $\Nf$ flipped spins in the ferromagnetic background~\cite{Lohani2019}. Because the Hamiltonian is invariant under global spin rotations about $z$, the classical helicity $\chi$ of such a state is a zero mode; in an eigenstate of fixed $\Nf$ it is conjugate to $\Nf$ and therefore completely undetermined. Only the relative helicity of two skyrmions is an observable in this setting, and it is this quantity that we measure.

\begin{figure}[t]
\centering\includegraphics[width=\columnwidth]{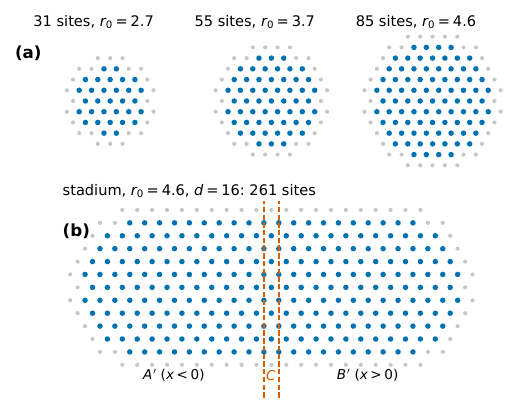}
\caption{Flake geometries. (a) Discs of 31, 55, and 85 sites used for a single skyrmion. (b) Stadium of half-width $r_0=4.6$ and length $d=16$ (261 sites) used for the free pair. Gray dots mark the first shell of the frozen ferromagnetic background. The dashed lines delimit the central column $C$ ($x=0$); the regions $A'$ ($x<0$) and $B'$ ($x>0$) are used for the number distribution $P(n_A)$ and the mutual information (Appendix~\ref{app:ent}).}
\label{fig:geo}
\end{figure}

We work on finite flakes embedded in a frozen ferromagnetic background: all spins outside the flake are fixed to $S^z=-1/2$ and act on the boundary spins through effective fields, $-J S_i^z S_j^z\to +(J/2)S_i^z$ for each missing bond. Three flake shapes are used (Fig.~\ref{fig:geo}): discs of radius $r_0=2.7$, $3.7$, and $4.6$ lattice spacings (31, 55, and 85 sites) for a single skyrmion, and \emph{stadiums}, the set of sites within $r_0$ of a segment of length $d$, for two skyrmions, with $r_0=3.7$, $4.6$, and $5.5$ and lengths up to $d=16$ (261 sites). Optionally, a weak local field $h_w e^{-r^2/2\sigma^2}S_i^z$ ($\sigma=0.8$) centered at each end of the segment pins the positions without acting on the helicity; the main results are obtained with $h_w=0$. The frozen background compresses a skyrmion that touches it, so every energy reported below is tested against the flake size.

\subsection{Exact diagonalization}
For flakes up to 31 sites we diagonalize $H$ exactly in the $\Nf$ sector (up to $2.6\times10^{6}$ states for $\Nf=7$); a matrix-free Lanczos solver extends this to $\sim10^{8}$ states. On the 31-site disc the $60^\circ$ rotation is a symmetry and each eigenstate carries an angular momentum $l_z\in\{0,\dots,5\}$. We reproduce the selection rule of Ref.~\cite{Lohani2019}, $l_z=\pm(\Nf-1)\bmod 6$ for the skyrmion and antiskyrmion ground states, the corresponding degeneracies, and the skyrmion--antiskyrmion tunnel splitting at $\Nf=7$ and $K=0.05$, $\Delta_t=0.0501J_1$, against the value $\approx0.05J_1$ quoted in that work. At $K=0.03$ the splitting is $\Delta_t=0.0447J_1$.

\subsection{DMRG}
Larger flakes are treated with DMRG~\cite{White1992} as implemented in ITensor~\cite{itensor}, with explicit U(1) symmetry, i.e., fixed $\Nf$. Sites are ordered column by column (increasing $x$, then $y$), so that the left half of a stadium is a contiguous prefix of the matrix-product state (MPS) and the cut between the two halves is a single bond. Bond dimensions up to $\chi=2000$ are used; energies are converged to $\sim10^{-3}J_1$ for the pair states, whose entanglement is larger than that of a single skyrmion (Appendix~\ref{app:conv}). The DMRG was validated against exact diagonalization on the 31-site disc ($\Nf=7$: $E-E_{\rm fm}=-0.88424J_1$ in both) and on a 29-site two-center flake ($\Nf=8$: energy, number distribution, and entanglement entropy agree to all printed digits). The initial state places $\Nf/2$ flipped spins around each of two chosen points. Since DMRG is a variational minimization and not a dynamics, the final positions can depend on the start when the energy landscape is shallow; we therefore run several starts and keep the lowest energy (Sec.~\ref{sec:free}).

\subsection{Observables}
Let $E(\Nf)$ be the ground-state energy of the $\Nf$ sector and $E_{\rm fm}$ that of the fully polarized flake. A single flipped spin forms a magnon whose band minimum in the infinite ferromagnet is $E^m_{\min}=3K-1/4$ for $J_2=0.5$; it is negative at $B=0$ because of the negative stiffness, which is irrelevant at fixed $\Nf$. The binding energy per flipped spin,
\begin{equation}
e(\Nf)=\frac{E(\Nf)-E_{\rm fm}-\Nf E^m_{\min}}{\Nf},
\end{equation}
is negative for a bound state~\cite{Lohani2019}. The pair interaction energy is
\begin{equation}
U=E(2\Nf)-2E(\Nf),
\label{eq:U}
\end{equation}
with both energies computed in the same flake, so that $E_{\rm fm}$ and the boundary fields cancel. To distinguish two skyrmions from one larger object we cut the stadium at its midpoint and compute the probability distribution $P(n_A)$ of the number of flipped spins in the left half, obtained from the charge blocks of the Schmidt decomposition at the cut, together with the column profile $n(x)=\sum_{i\in x}(\langle S^z_i\rangle+1/2)$. Two skyrmions of $\Nf$ flipped spins each give a peak of $P(n_A)$ at $\Nf$ and two lobes in $n(x)$; a single delocalized object gives a broad or bimodal $P(n_A)$. Finally, the in-plane correlations $\langle S^+_iS^-_j\rangle$ between the walls of the two skyrmions, weighted by the vortex phase of each, measure the coherence of the relative helicity (Sec.~\ref{sec:hel}).

\section{A single quantum skyrmion has a natural size}
\label{sec:single}
\begin{figure}[t]
\centering\includegraphics[width=\columnwidth]{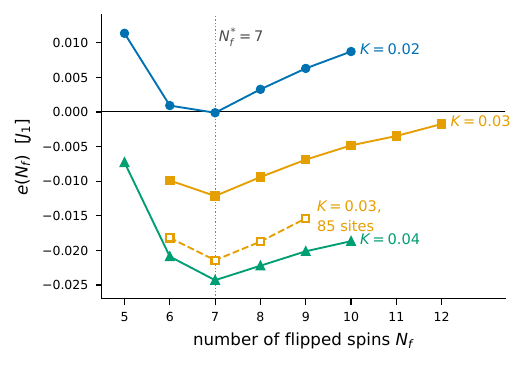}
\caption{Binding energy per flipped spin, $e(\Nf)$, of a single quantum skyrmion (DMRG, $\chi=1000$--$1200$) on the 55-site disc for $K=0.02$, $0.03$, and $0.04$ (filled symbols), and on the 85-site disc for $K=0.03$ (open symbols). The minimum at $\Nf^*=7$ (dotted line) defines the natural size; its position is independent of $K$ and of the flake, while its depth is not.}
\label{fig:eNf}
\end{figure}

Figure~\ref{fig:eNf} shows $e(\Nf)$ for $K=0.02$, $0.03$, and $0.04$ on the 55-site disc, and for $K=0.03$ on the 85-site disc. In all cases $e(\Nf)$ has a minimum at
\begin{equation}
\Nf^{*}=7 ,
\end{equation}
and rises on both sides: $e=-0.0099$, $-0.0121$, $-0.0094$, and $-0.0068$ for $\Nf=6$, 7, 8, and 9 at $K=0.03$ on the 55-site disc, and $-0.0182$, $-0.0214$, $-0.0187$, and $-0.0154$ on the 85-site disc. The absolute value of $e$ depends on the flake, since the frozen background compresses the skyrmion and $e(7)$ decreases by $0.009$ when the disc radius grows from $3.7$ to $4.6$, but the position of the minimum does not. At $K=0.02$ the skyrmion is only marginally bound on the 55-site disc, $e(7)=-0.0001$; the binding deepens with $K$, $e(7)=-0.0243$ at $K=0.04$. The minimum at $\Nf^*=7$ is the same for the three anisotropies, so in this range the natural size of the quantum skyrmion is set by the frustration ratio $J_2/J_1$, while $K$ controls only how strongly it is bound.

The existence of this minimum is the ingredient on which everything that follows depends. For a larger anisotropy there is none: at $K=0.20$, $e(\Nf)=+0.046$, $-0.034$, $-0.116$, and $-0.164$ for $\Nf=2,\dots,5$, and it keeps decreasing, i.e., the flipped spins prefer to form a single large domain (the Ising-like regime of Ref.~\cite{Lohani2019}). As shown next, in that regime two skyrmions cannot coexist.

\section{Below the natural size, two skyrmions coalesce}
\label{sec:coal}
Consider $2\Nf$ flipped spins in a stadium. Since the Zeeman energy is constant in the sector, the competition between two skyrmions of $\Nf$ flipped spins and one object of $2\Nf$ is decided by the convexity of $E(\Nf)$ alone: two objects win only if $E(2\Nf)>2E(\Nf)$, i.e., if $e(2\Nf)>e(\Nf)$, which requires $e$ to have a minimum below $2\Nf$. Where $e(\Nf)$ decreases monotonically, two small skyrmions placed at any distance must merge.

\begin{figure}[t]
\centering\includegraphics[width=\columnwidth]{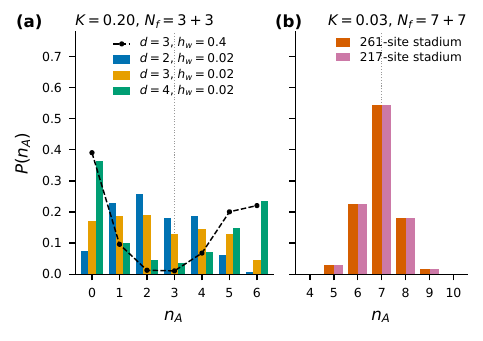}
\caption{Distribution $P(n_A)$ of the number of flipped spins in the left half of a stadium. (a) $K=0.20$, $\Nf=3+3$ skyrmions placed at $d=2$, 3, and 4 with weak wells ($h_w=0.02$): $P(n_A)$ is broad and has no peak at $n_A=3$; strong wells ($h_w=0.4$, dashed line) make it bimodal. The two skyrmions have coalesced. (b) $K=0.03$, $\Nf=7+7$, free pair ($h_w=0$) on the 217- and 261-site stadiums: a sharp peak at $n_A=7$ identifies two skyrmions.}
\label{fig:coal}
\end{figure}

Figure~\ref{fig:coal}(a) shows this for $K=0.20$ and $\Nf=3+3$ on stadiums of half-width $r_0=2.7$ (Lanczos, up to $4.5\times10^{7}$ states); the $\Nf=3$ state is a skyrmion on the 31-site disc, with the $l_z=2,4$ doublet required by the selection rule. For center separations $d=2$, 3, and 4 the interaction energy is $U\simeq-0.96$, $-0.95$, and $-0.94J_1$, an order of magnitude larger than any skyrmion binding scale, and $P(n_A)$ is broad with little weight at $n_A=3$: the six flipped spins form a single object delocalized between the two ends. Pinning wells do not help. Increasing $h_w$ from $0.02$ to $0.4$ makes $P(n_A)$ bimodal [all six spins in one well or in the other, dashed line in Fig.~\ref{fig:coal}(a)], because a six-spin object still fits in a well and the gain from coalescence ($\sim J_1$) exceeds the energy difference between the two well configurations ($\sim0.5h_w$). The same happens for $\Nf=4+4$ at $K=0.10$. We conclude that quantum skyrmions smaller than the natural size do not exist as pairs; the pair problem is only defined at $\Nf=\Nf^*$.

\section{At the natural size, two skyrmions form a bound molecule}
\label{sec:mol}

\subsection{Pinned pair: shape of the interaction}
\label{sec:pinned}
\begin{figure}[t]
\centering\includegraphics[width=\columnwidth]{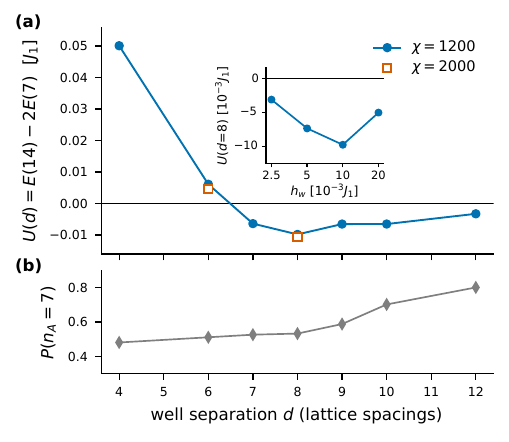}
\caption{Pinned pair at $K=0.03$ ($h_w=0.01$, $r_0=3.7$ stadiums). (a) Interaction energy $U(d)=E(14)-2E(7)$ as a function of the well separation $d$; open squares, $\chi=2000$. Inset: $U$ at $d=8$ as a function of the pinning strength $h_w$. (b) Weight $P(n_A=7)$ of the $7+7$ configuration, showing that the pair remains two objects at all $d$.}
\label{fig:Ud}
\end{figure}

We first pin two $\Nf=7$ skyrmions with weak wells ($h_w=0.01$) at separations $d=4$--$12$ on $r_0=3.7$ stadiums (91--163 sites) at $K=0.03$, and compute $U(d)$ from Eq.~(\ref{eq:U}) with the single-skyrmion reference in the same stadium (Fig.~\ref{fig:Ud}). In every case $P(n_A)$ peaks at $n_A=7$ (weight $0.48$--$0.80$), so the pair never coalesces. $U(d)$ is strongly repulsive at short distance ($+0.050$ at $d=4$), changes sign between $d=6$ and 7, is attractive for $d\ge7$ with a minimum of $-0.010$ near $d=8$, and decays toward zero ($-0.003$ at $d=12$). Increasing $\chi$ from 1200 to 2000 changes $U$ by at most $10^{-3}$. The depth of the well, however, depends on the pinning strength in a nonmonotonic way [inset of Fig.~\ref{fig:Ud}(a): $U(8)=-0.003$, $-0.007$, $-0.010$, and $-0.005$ for $h_w=0.0025$, $0.005$, $0.01$, and $0.02$], because the single-skyrmion reference delocalizes between the two wells when they are weak and is compressed by them when they are strong. The pinned curve therefore gives the shape of the interaction, a short-range repulsion followed by an attraction beyond $d\simeq6$, but not a reliable binding energy. For that we remove the wells.

\subsection{Free pair: binding without pinning}
\label{sec:free}
\begin{figure}[t]
\centering\includegraphics[width=\columnwidth]{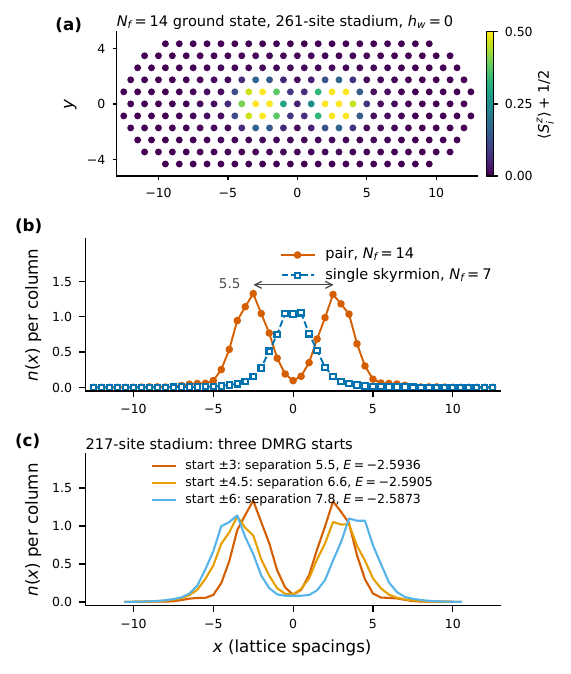}
\caption{Free pair ($h_w=0$, $K=0.03$). (a) Local density of flipped spins, $\langle S^z_i\rangle+1/2$, in the $\Nf=14$ ground state of the 261-site stadium. (b) Column profile $n(x)$ of the same state and of the single-skyrmion reference ($\Nf=7$) in the same stadium; the two lobes of the pair are $5.5$ lattice spacings apart. (c) Three DMRG starts on the 217-site stadium end at different separations with monotonically increasing energy (in units of $J_1$, relative to $E_{\rm fm}$); the lowest, at separation $5.5$, is the ground state.}
\label{fig:free}
\end{figure}

With $h_w=0$ the skyrmions go where the energy takes them. Figure~\ref{fig:free}(a) shows the map of $\langle S^z_i\rangle+1/2$ for the $\Nf=14$ ground state on the 261-site stadium ($r_0=4.6$, $d=16$), and Fig.~\ref{fig:free}(b) the corresponding column profile $n(x)$: two lobes of $6.93$ and $7.07$ flipped spins (in the halves $x<0$ and $x\ge0$) centered at $x=-2.75$ and $+2.79$, separated by $5.5$ lattice spacings, with $n(0)=0.09$ in between, and $P(n_A)$ peaked at 7 with weight $0.55$ [Fig.~\ref{fig:coal}(b)]. The single-skyrmion reference in the same stadium sits at the center as one lobe of seven spins. The energies give
\begin{equation}
U_{\rm free}=E(14)-2E(7)=-0.0066J_1
\label{eq:Ufree}
\end{equation}
at $K=0.03$ and $\chi=1200$, i.e., the pair is bound by about $6\%$ of the binding energy of one skyrmion ($\Nf^*|e(7)|\simeq0.1$--$0.15J_1$ on these flakes).

Two artifacts had to be controlled to obtain this number. The first is the existence of local minima. DMRG converges to the nearest minimum of the shallow intermolecular potential: on the 217-site stadium, starts with the two skyrmions at $\pm3$, $\pm4.5$, and $\pm6$ end at separations $5.5$, $6.6$, and $7.8$ with energies $-2.5936$, $-2.5905$, and $-2.5873$ [Fig.~\ref{fig:free}(c)], which increase monotonically with the separation. The equilibrium separation is thus $5.5$, and we always take the lowest of several starts. The second is the effect of the rounded ends of the stadium. A single skyrmion that stops near an end is compressed and its energy is $\sim4\times10^{-3}$ too high, which would spuriously double the binding; the reference must be started at the center, where its energy is independent of the stadium length ($-1.29418$ and $-1.29421$ for $d=12$ and 16 at $\chi=1000$). With these controls, $U_{\rm free}$ is $-0.0052$ on the 217-site stadium and $-0.0044$ ($\chi=1000$) or $-0.0066$ ($\chi=1200$) on the 261-site one. The remaining drift is the bond-dimension convergence of the more entangled 14-spin state, which lowers $U$ further (Appendix~\ref{app:conv}); we quote $U_{\rm free}=-0.0066\pm0.002J_1$.

The width of the stadium does not change the picture. On the narrow $r_0=3.7$ stadium ($d=12$, 163 sites) the free pair binds at a separation of $5.8$, and on the wide $r_0=5.5$ stadium ($d=8$, 213 sites) at $5.4$--$5.6$. We do not quote binding energies from these two geometries: in the narrow one the frozen background compresses both the pair and the reference, and in the wide one the segment is too short for the two skyrmions, which sit at $\pm2.8$, to stay clear of the rounded ends that the centered reference does not feel. The length test ($d=12$ versus $16$ at $r_0=4.6$) is the one in which neither state touches the ends, and it is on that pair of stadiums that Eq.~(\ref{eq:Ufree}) rests.

\subsection{Dependence on the anisotropy}
Repeating the free-pair calculation at $K=0.02$ and $0.04$ on the 217-site stadium (lowest of the starts, centered reference, $\chi=1000$) gives $U_{\rm free}=+0.002$, $-0.005$, and $-0.006J_1$ for $K=0.02$, $0.03$, and $0.04$, with equilibrium separations of $5.6$, $5.5$, and $5.5$. At $K=0.02$, where the single skyrmion is itself only marginally bound (Fig.~\ref{fig:eNf}), the pair is unbound within our resolution: the anisotropy that binds the skyrmion is the one that binds the molecule. The equilibrium separation, on the other hand, is insensitive to $K$, like $\Nf^*$; both are set by the frustration.

\section{Coherent magnon exchange between the two skyrmions}
\label{sec:hel}
Classically, the interaction between two skyrmions in a frustrated magnet contains a term $W(d)\cos(\chi_A-\chi_B)$ that depends on their relative helicity~\cite{Leonov2015}. In the quantum problem $\chi$ is conjugate to $\Nf$, so $e^{i(\chi_A-\chi_B)}$ is the operator that transfers one flipped spin from skyrmion $B$ to skyrmion $A$, and $\cos(\chi_A-\chi_B)$ is a coherent exchange of magnons between the two. Its signature should therefore be visible in two places: as fluctuations of the number of flipped spins on either side of the midpoint cut beyond those produced by the tails of two independent skyrmions, and as coherent in-plane correlations $\langle S^+_iS^-_j\rangle$ between the wall of $A$ and the wall of $B$.

\begin{figure}[t]
\centering\includegraphics[width=\columnwidth]{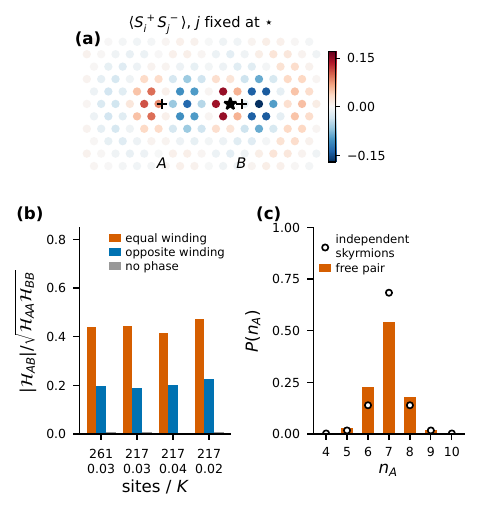}
\caption{Coherent magnon exchange in the free pair. (a) In-plane correlation $\langle S^+_iS^-_j\rangle$ between every site $i$ and the site $j$ (star) on the wall of skyrmion $B$ that faces skyrmion $A$ ($K=0.03$, 261 sites); crosses mark the two lobe centers. The correlation reaches $0.16$ across the gap and follows the vortex pattern of skyrmion $A$. (b) Normalized sum $|\mathcal H_{AB}|/\sqrt{\mathcal H_{AA}\mathcal H_{BB}}$, Eq.~(\ref{eq:HAB}), for the four free pairs (labeled by stadium size and $K$; wall window $0.8$--$2.2$), with the vortex phases of equal winding, of opposite winding, and without phases. (c) Distribution $P(n_A)$ of the free pair on the 261-site stadium ($K=0.03$) compared with that of two independent skyrmions at the same positions, obtained by convolving the tail distribution of a single skyrmion placed at $x=+2.75$ with its mirror image; the variances are $0.61$ and $0.47$.}
\label{fig:corr}
\end{figure}

\paragraph*{Number fluctuations.}
For the free pair on the 261-site stadium the variance of $n_A$ across the midpoint cut is ${\rm var}(n_A)=0.61$. To estimate the contribution of the tails alone we place a single skyrmion at $x=+2.75$, the position of one member of the pair, in the same stadium, read off the distribution of its flipped spins that cross the cut, and convolve it with its mirror image. The resulting distribution for two independent skyrmions is compared with $P(n_A)$ of the pair in Fig.~\ref{fig:corr}(c): the pair has less weight at $n_A=7$ ($0.55$ against $0.68$) and more at $n_A=6$ and $8$, and its variance, $0.61$, exceeds the independent value, $0.47$, by $0.14$. This excess is the exchange channel. The mutual information between the two halves of the stadium shows the same excess, about $0.5$~nats beyond that of two independent skyrmions (Appendix~\ref{app:ent}).

\paragraph*{In-plane correlations.}
The in-plane correlations between the two walls, defined as the sites at distance $0.8$--$2.2$ from each lobe center, reach $|\langle S^+_iS^-_j\rangle|=0.16$ for facing sites $5.5$ lattice spacings apart, out of a maximum of $0.25$ [Fig.~\ref{fig:corr}(a)]. In the classical texture of a skyrmion of winding $\sigma$ and helicity $\chi$, the in-plane component on the wall is $S^+_i\propto e^{i(\sigma\varphi_i+\chi)}$, with $\varphi_i$ the polar angle of site $i$ around the center, so that for two skyrmions of equal winding $\langle S^+_iS^-_j\rangle\propto e^{i(\varphi_i-\varphi_j)}\langle e^{i(\chi_A-\chi_B)}\rangle$. The sum
\begin{equation}
\mathcal H_{AB}=\sum_{i\in A,\,j\in B}e^{-i(\varphi_i-\varphi_j)}\langle S^+_iS^-_j\rangle,
\label{eq:HAB}
\end{equation}
normalized by $\sqrt{\mathcal H_{AA}\mathcal H_{BB}}$, is therefore an estimate of the coherence of the relative helicity, $|\langle e^{i(\chi_A-\chi_B)}\rangle|$, which would be 1 for a fully locked pair and 0 for independent helicities. For the four free pairs we find $0.441$ (261 sites) and $0.443$ (217 sites) at $K=0.03$, $0.416$ at $K=0.04$, and $0.471$ at $K=0.02$ [Fig.~\ref{fig:corr}(b)], while the same sum with opposite winding phases, $e^{-i(\varphi_i+\varphi_j)}$, gives $0.19$--$0.22$ and the unweighted sum is $\lesssim0.01$, as it must be for vortex textures. The absolute value depends on the wall window: for the 261-site pair, windows between $0.5$--$2.5$ and $1.0$--$2.0$ give $0.52$--$0.54$ for equal winding and $0.31$--$0.37$ for opposite winding. A least-squares fit of the wall--wall correlation matrix to $a\cos(\varphi_i-\varphi_j)+b\cos(\varphi_i+\varphi_j)+c$ (with the corresponding sine terms, which come out negligible) gives $a=0.077$, $b=0.048$, and $c\simeq0$, and accounts for $84\%$ of the variance of the data. Two conclusions follow and do not depend on the window. The in-plane correlation between the two walls is dominated by the equal-winding harmonic $e^{i(\varphi_i-\varphi_j)}$, which is the quantized form of the $W\cos(\chi_A-\chi_B)$ coupling: the relative helicity of the two skyrmions is correlated with a coherence of about one half. And the opposite-winding harmonic is present with a smaller but finite amplitude, which is what one expects if each constituent is, like the isolated $\Nf=7$ skyrmion~\cite{Lohani2019,Liscak2026}, a superposition of skyrmion and antiskyrmion. The pair thus lives in the four-state manifold $\{\sigma_A\sigma_B\}$, and the correlations measured here are those of the DMRG ground state within it. These values are the same, within a few percent, for the two stadium lengths and the three anisotropies.

\section{Discussion}
\label{sec:disc}
Three energy scales characterize the molecule at $K=0.03$: the binding energy of one skyrmion, $\Nf^*|e(7)|\simeq0.1$--$0.15J_1$; the skyrmion--antiskyrmion tunnel splitting of one skyrmion, $\Delta_t=0.045J_1$; and the binding energy of the pair, $|U_{\rm free}|\simeq0.007J_1$. The molecule is therefore a weakly bound object, in the sense of a van der Waals dimer rather than a covalent one: each skyrmion keeps its seven flipped spins and its identity, and the pair is held by a shallow well. For a material with $J_1\sim10$~meV the binding would be of the order of $1$~K, which sets the temperature range in which the molecule could be observed.

The lowest excitations of the free pair on the 217-site stadium, obtained with the penalty method at $\chi=800$, lie $3.2$, $5.5$, and $7.7\times10^{-3}J_1$ above the ground state. These gaps are of the same order as the truncation error of the pair energy at this bond dimension, so they should be read as an indication rather than a measurement: they suggest that the molecule has soft modes of relative and center-of-mass motion with quanta of a few $10^{-3}J_1$, comparable to the depth of the well, i.e., one or two bound vibrational levels. A bond dimension sufficient to resolve this spectrum on 217 sites exceeds the memory available to us.

The physics behind the binding is the convexity of $E(\Nf)$, a quantum many-body property with no classical counterpart: a classical skyrmion has a continuous size set by the balance of exchange and anisotropy, whereas the quantum one has an integer number of flipped spins and a discrete $e(\Nf)$ with a minimum. Below the minimum, merging always wins; at the minimum, two objects are stable and the residual interaction decides their arrangement. This is the quantum analog of the statement that two classical skyrmions in a frustrated magnet attract or repel depending on their relative helicity~\cite{Leonov2015,Zhang2017,Capic2020}, with one difference: in an eigenstate of fixed $\Nf$ the helicity of each skyrmion does not exist, only the relative one does, and it is this relative helicity that the interaction correlates.

For the skyrmion-qubit program this has a direct consequence. The coupling between two helicity qubits does not have to be engineered through a bus: it is contained in the pair interaction, and the molecule assembles itself at the distance where that coupling is largest. The strength of the coupling, $W$, should be compared with $\Delta_t$. We could not extract it from the spectrum, because in a large flake the lowest excitations of both one skyrmion and the pair are soft modes of position, with gaps of a few $10^{-3}J_1$, below which the helicity manifold is buried. The excess number fluctuations, the mutual information, and the wall--wall correlations of Sec.~\ref{sec:hel} are the quantities accessible to DMRG. A clean separation of the entanglement between the two helicity degrees of freedom from the spatial correlations of the tails is left for future work, as is the spectrum of the four-state manifold $\{\sigma_A\sigma_B\}$, which would require a symmetry-resolved method.

The quantum skyrmion molecule should be distinguished from the biskyrmion of centrosymmetric dipolar magnets~\cite{Yu2014,Gobel2019}. The biskyrmion is a pair of partially overlapping skyrmions with a helicity difference of $\pi$, held together by the dipolar interaction, and it is a single object with topological charge $2$. The molecule studied here is the opposite case in each respect: the two skyrmions do not overlap (the density between them drops to $0.09$ flipped spins per column), the dominant wall--wall correlation is the equal-winding harmonic rather than the $\pi$-shifted one, and the binding comes from the exchange of magnons in a model without dipolar interaction. The bound pairs of classical skyrmions reported in frustrated and dipolar films~\cite{Zhang2017,Lee2016} are closer in spirit, but there the helicity of each skyrmion is a classical variable fixed by the dipolar energy; in the molecule it is not defined, and only the relative helicity carries information.

We have deliberately worked at a single frustration ratio, $J_2/J_1=0.5$, where the selection rule of Ref.~\cite{Lohani2019} holds and the skyrmion is cleanly identified. At larger $J_2$ ($0.6$--$0.7$) the ground states of the few-flipped-spin sectors on the 31-site disc carry different angular momenta and are not the same objects; whether those also pair is an open question. The dipolar skyrmions of thin films are not the place to look for this physics: there the dipolar interaction breaks the spin-rotation symmetry, the helicity is pinned to the Bloch value by the magnetostatic energy, and the skyrmion contains $10^{4}$--$10^{5}$ spins, which suppresses any helicity coherence exponentially. The frustrated magnet with a few-spin skyrmion is the regime in which the question makes sense; candidate materials are the centrosymmetric frustrated magnets in which skyrmion lattices have been observed~\cite{Kurumaji2019,Petrovic2025} and van der Waals magnets with competing exchange.

\section{Conclusions}
In the frustrated spin-$1/2$ triangular magnet of Ref.~\cite{Lohani2019}, a quantum skyrmion has a natural size of $\Nf^*=7$ flipped spins, set by the frustration and independent of the anisotropy for $K=0.02$--$0.04$. Two skyrmions smaller than this size coalesce at any distance, and no pinning can keep them apart. Two skyrmions of natural size form a bound molecule, with binding energy $\simeq0.007J_1$ at $K=0.03$--$0.04$ and equilibrium separation of $5.5$ lattice spacings, held by the coherent exchange of magnons between them. This exchange appears as number fluctuations between the two skyrmions in excess of those of independent objects and as in-plane correlations between the two walls that follow the vortex pattern of the textures and correlate their relative helicity. The molecule is a candidate two-qubit element for skyrmion helicity qubits in the same layer.

\begin{acknowledgments}
The calculations were carried out on the author's own equipment; this work received no external funding. The DMRG calculations used the ITensor library~\cite{itensor}.
\end{acknowledgments}

\section*{Data availability}
The exact-diagonalization, Lanczos, and DMRG outputs that support every figure and table of this paper (energies, local magnetizations, number distributions, entanglement entropies, the in-plane correlation matrix of the free pair, and the geometry files), together with all scripts and a script that regenerates the figures, are openly available in the Zenodo repository~\cite{zenodo}. The matrix-product-state wavefunctions (13~GB) are available from the author upon reasonable request.

\appendix
\section{Convergence tests}
\label{app:conv}
Table~\ref{tab:conv} collects the checks on the free-pair binding energy at $K=0.03$. The single-skyrmion energy is converged in $\chi$ to $2\times10^{-4}$ and independent of the stadium length once the skyrmion is started at the center; the pair energy, more entangled, still decreases by $2.6\times10^{-3}$ between $\chi=1000$ and $1200$ on the 261-site stadium, so the quoted $U_{\rm free}$ is an upper bound (less binding), and the trend with $\chi$ points to a slightly deeper well. A larger $\chi$ for 261 sites exceeds the memory of our machine. In the first row of the table $E(7)$ was computed at $\chi=1000$ and $E(14)$ at $\chi=1200$; since $E(7)$ changes by $2\times10^{-4}$ between the two bond dimensions, this mixing affects $U_{\rm free}$ by at most $4\times10^{-4}$, within the quoted error. The dependence on the starting separation (local minima) and on the position of the reference (ends of the stadium) is discussed in Sec.~\ref{sec:free}; the pinned-pair curve at $\chi=1200$ and $2000$ agrees to $10^{-3}$ (Fig.~\ref{fig:Ud}).
\begin{table}[h]
\caption{Free pair, $K=0.03$, $r_0=4.6$: energies (in units of $J_1$, relative to $E_{\rm fm}$) of the centered single skyrmion, $E(7)$, of the pair started at $\pm3$, $E(14)$, and $U_{\rm free}$, for two stadium lengths and bond dimensions.}
\label{tab:conv}
\begin{ruledtabular}
\begin{tabular}{lccccc}
sites & $d$ & $\chi$ & $E(7)$ & $E(14)$ & $U_{\rm free}$\\
\hline
217 & 12 & 1000/1200 & $-1.29418$ & $-2.59359$ & $-0.0052$\\
261 & 16 & 1000 & $-1.29421$ & $-2.59284$ & $-0.0044$\\
261 & 16 & 1200 & $-1.29443$ & $-2.59547$ & $-0.0066$\\
\end{tabular}
\end{ruledtabular}
\end{table}

\section{Entanglement across the cut}
\label{app:ent}
With the left half $A$ of the stadium contiguous in the MPS, the Schmidt decomposition at the cut gives the entanglement entropy $S_A$ and, because the U(1) charge of each Schmidt block is the number of flipped spins on the left, the distribution $P(n_A)$ without further work. For the mutual information between the two skyrmions we take $A'=\{x<0\}$, $B'=\{x>0\}$, and the central column $C=\{x=0\}$ (five sites); since the state is pure, $S(A'\cup B')=S_C$, which we compute from the explicit $2^5\times2^5$ reduced density matrix of $C$, so that $I(A'{:}B')=S_{A'}+S_{B'}-S_C$ requires only two bond cuts and one small block. Both $S_A$ and $I(A'{:}B')$ contain the contribution of the tails of each skyrmion crossing the cut, which we estimate with a single skyrmion placed at the position of one member of the pair ($x=+2.75$): $S_{A'}=0.82$, $S_{B'}=1.18$, $S_C=0.54$, and $I=1.45$~nats, against $1.91$, $1.90$, $0.39$, and $3.42$~nats for the pair. If the two skyrmions were independent the mutual information would be additive, $\simeq2.9$~nats; the excess of $\sim0.5$~nats is the correlation with the partner. The subtraction is approximate because the two tails overlap in $C$.

\bibliographystyle{apsrev4-2}
\bibliography{refs}
\end{document}